\documentclass[aps,prl,reprint,longbibliography]{revtex4-1}
\usepackage{appendix}
\usepackage{bm}
\usepackage{enumerate}
\usepackage{graphicx}% Include figure files
\usepackage{dcolumn}% Align table columns on decimal point
\usepackage{bm}% bold math
\usepackage{amssymb}
\usepackage{amsmath}
\usepackage{booktabs}
\usepackage{braket}
\usepackage{float}
\usepackage[caption = false]{subfig}

\usepackage{tikz}

\newcommand{\ee}{{\rm e}}
\newcommand{\ii}{{\rm i}}

\tikzset{every picture/.style={line width=0.75pt}} %set default line width to 0.75pt        

\begin{document}

\preprint{APS/123-QED}

\title{Correlations of the Current Density in Many-Body Landau Level States}% Force line breaks with \\

\author{Daniel Spasic-Mlacak  and Nigel R. Cooper}

\affiliation{
T.C.M Group, Cavendish Laboratory, University of Cambridge, J.J. Thompson Avenue, Cambridge CB3 0HE, United Kingdom \looseness=-1}
\date{\today}

\begin{abstract}
Motivated by recent advances in quantum gas microscopy, we investigate correlation functions of the current density in many-body Landau Level states, such as the Laughlin state of the fractional quantum Hall effect. For states fully in the lowest Landau level, we present an exact relationship which shows that all correlation functions involving the current density are directly related to correlation functions of the number density. We calculate perturbative corrections to this relationship arising from inter-particle interactions, and show that this provides a method by which to extract the system's interaction energy. Finally, we demonstrate the applicability of our results also to lattice systems. 
\end{abstract}

\maketitle

%\tableof dfcontents

Recently, significant progress has been made towards the realisation of fractional quantum Hall (FQH) and other many-body Landau Level (LL) states in quantum simulators. Many  ways to generate effective magnetic fields have been proposed for cold-atom and photonic systems~\cite{cds,topophotonics},
and have been successfully realised experimentally~\cite{jotzu,aidelsburgerEffectiveBfield,aidelsburger2013realization,clark2020observation,weber2022experimentally}, including for strongly interacting bosons in flat Chern bands~\cite{leonard2023realization,FCIReview}. Recent experiments have also studied the evolution of Bose-Einstein condensates near the lowest Landau Level (LLL) of rotating gases \cite{RotatingBoson}, and of strongly-correlated few-fermion systems~\cite{lunt2024realizationlaughlinstaterapidly}. These quantum simulators offer the potential to study the FQH effect in new settings, such as investigating bosons rather than fermions and different types of interactions between particles, as well as to probe and detect properties of the FQH states in novel ways~\cite{halperinjain-cooper}. 

One of the most powerful methods of extracting information from quantum simulations is through measurements of multi-point correlation functions~\cite{qgm,leonard2023realization,lunt2024realizationlaughlinstaterapidly}, as has been applied to other systems such as atomic superfluids~\cite{ExpCorr} and the Fermi-Hubbard model~\cite{cheuk2016observation, parsons2016site}.
Motivated by recent advances in quantum gas microscopy which allow for the measurement of local {\it current} operators in optical  lattices \cite{ExpCurrent}, we investigate current density correlation functions of FQH states (and, more broadly, any generic many-body LL states) to gain insight into how the properties of such states can be probed by measurements of these correlations. 

In this paper, we report on an exact relationship that holds in the LLL between correlation functions of the current density and correlation functions of the number density. We show that deviations from this relationship arise from inter-particle interactions, and hence that measurements of the current density and number density correlators can be used as a way to determine the interaction energy of the many-body system experimentally. Finally, we discuss how our findings can be related to lattice settings, and show that a discrete version of the LLL relationship between correlators still holds for single-particle states in the Harper-Hofstadter model at low flux.  

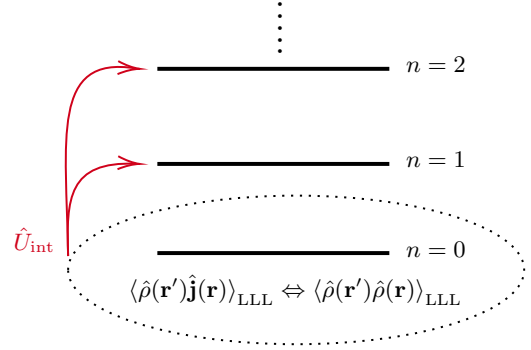
\begin{figure}
\centering

\tikzset{every picture/.style={line width=0.75pt}} %set default line width to 0.75pt        

\begin{tikzpicture}[x=0.75pt,y=0.75pt,yscale=-1,xscale=1]

%Straight Lines [id:da42034836401465125] 
\draw [line width=1.5]    (184,211) -- (301,211) ;
%Straight Lines [id:da37954341566987126] 
\draw [line width=1.5]    (184,166) -- (301,166) ;
%Straight Lines [id:da9750684210047667] 
\draw [line width=1.5]    (184,118) -- (301,118) ;
%Shape: Ellipse [id:dp9565818713441566] 
\draw  [dash pattern={on 0.84pt off 2.51pt}] (139,219.5) .. controls (139,198.79) and (190.49,182) .. (254,182) .. controls (317.51,182) and (369,198.79) .. (369,219.5) .. controls (369,240.21) and (317.51,257) .. (254,257) .. controls (190.49,257) and (139,240.21) .. (139,219.5) -- cycle ;
%Curve Lines [id:da2288495596316925] 
\draw [color={rgb, 255:red, 208; green, 2; blue, 27 }  ,draw opacity=1 ]   (139,212.5) .. controls (138.01,183.44) and (138.97,164.57) .. (173,165.92) ;
\draw [shift={(175,166)}, rotate = 183.18] [color={rgb, 255:red, 208; green, 2; blue, 27 }  ,draw opacity=1 ][line width=0.75]    (10.93,-3.29) .. controls (6.95,-1.4) and (3.31,-0.3) .. (0,0) .. controls (3.31,0.3) and (6.95,1.4) .. (10.93,3.29)   ;
%Curve Lines [id:da6602357017232784] 
\draw [color={rgb, 255:red, 208; green, 2; blue, 27 }  ,draw opacity=1 ]   (139,212.5) .. controls (138.01,153.59) and (130.16,110.86) .. (173,117.78) ;
\draw [shift={(175,118)}, rotate = 183.18] [color={rgb, 255:red, 208; green, 2; blue, 27 }  ,draw opacity=1 ][line width=0.75]    (10.93,-3.29) .. controls (6.95,-1.4) and (3.31,-0.3) .. (0,0) .. controls (3.31,0.3) and (6.95,1.4) .. (10.93,3.29)   ;

% Text Node
\draw (168,220.4) node [anchor=north west][inner sep=0.75pt]    {$\braket{\hat{\rho}(\textbf{r}^{\prime})\hat{\textbf{j}}(\textbf{r})}_{\rm LLL}\Leftrightarrow\braket{\hat{\rho}(\textbf{r}^{\prime})\hat{\rho}(\textbf{r})}_{\rm LLL}$};
% Text Node
\draw (308,158.4) node [anchor=north west][inner sep=0.75pt]    {$n=1$};
% Text Node
\draw (308,204) node [anchor=north west][inner sep=0.75pt]    {$n=0$};
% Text Node
   \draw (308,110.4) node [anchor=north west][inner sep=0.75pt]    {$n=2$};
% Text Node
\draw (110,196.4) node [anchor=north west][inner sep=0.75pt]  [color={rgb, 255:red, 208; green, 2; blue, 27 }  ,opacity=1 ]  {$\hat{U}_{\text{int}}$};
% Text Node
\draw (242.5,105) node [anchor=north west][inner sep=0.75pt]   [align=left] {$\boldsymbol{\cdot}$};
% Text Node
\draw (242.5,99) node [anchor=north west][inner sep=0.75pt]   [align=left] {$\boldsymbol{\cdot}$};
% Text Node
\draw (242.5,93) node [anchor=north west][inner sep=0.75pt]   [align=left] {$\boldsymbol{\cdot}$};
% Text Node
\draw (242.5,87) node [anchor=north west][inner sep=0.75pt]   [align=left] {$\boldsymbol{\cdot}$};
% Text Node
\draw (242.5,81) node [anchor=north west][inner sep=0.75pt]   [align=left] {$\boldsymbol{\cdot}$};

\end{tikzpicture}

\caption{In the LLL, there is a direct correspondence between current density and number density correlators. When contact interactions $\hat{U}_{\text{int}}$ are turned on, the LLL states are perturbatively mixed with higher LL states, causing a deviation to the correspondence between correlators. }
\label{Figure1}
\end{figure}

We consider particles moving in two dimensions and subjected to gauge fields, such that they are well described by particles of charge $e$ and mass $M$ acted on by a perpendicular magnetic field $B$. The single-particle energy levels are the discrete LLs with energy difference $\hbar\omega_{c}$ between each level, where $\omega_{c}={e B}/{M}$ is the cyclotron frequency. 
For the symmetric gauge, with the vector potential $A_{\alpha}(\textbf{r}) = \frac{B}{2}\epsilon_{\alpha\beta}r_{\beta}$ where $\alpha,\beta$ are co-ordinate indices and $\epsilon_{\alpha\beta}$ is the two-dimensional Levi-Civita symbol, the single particle states have definite angular momentum $m=0,1,2,\ldots$ The position representations of the single particle wavefunctions in the LLL are  
\begin{equation}
\label{LLLstate}
\braket{\textbf{r} | m} = \frac{1}{\sqrt{2\pi l_{B}^{2}  m!}} \left(\frac{r}{\sqrt{2}l_{B}}\right)^{m} \exp{\left(-\frac{r^{2}}{4l_{B}^{2}}\right)} \ee^{{\rm i}m\theta},  
\end{equation}
where $l_{B}=\sqrt{\frac{\hbar }{e B}}$ is the magnetic length, and $\textbf{r}=(r,\theta)$ is the position in polar co-ordinates.

We use these single-particle states to form a generic first-quantized $N$-body state in the LLL as
\begin{equation}
\label{LLLmainstate}
\ket{\psi_{\rm LLL}}=\sum_{\{m_{i}\}} c_{m_{1}m_{2}...m_{N}}\ket{m_{1}}\ket{m_{2}}...\ket{m_{N}},
\end{equation}
where  $|m_i\rangle$ denotes the single particle state for particle $i$, and the notation $\sum_{\{m_{i}\}}$ indicates $N$ sums over $m_{1},...,m_{N}$. The complex coefficients $c_{m_{1}m_{2}...m_{N}}$ are symmetric (antisymmetric) for bosons (fermions) under particle exchange. We are interested in calculating correlation functions of number density, $\hat{\rho}({\textbf{r}})$ and current density, $ \hat{\textbf{j}}(\textbf{r})$,  defined as 
\begin{eqnarray}
%\begin{align*}
\label{densitycurrent}
\hat{\rho}(\textbf{r}) & = & \sum_{i=1}^{N}\hat{\rho}_{i}(\textbf{r})=\sum_{i=1}^{N}\delta(\textbf{r}-\hat{\textbf{r}}_{i})
, \\ \hat{\textbf{j}}(\textbf{r}) & = & \sum_{i=1}^{N}\hat{\textbf{j}}_{i}(\textbf{r})=\sum_{i=1}^{N}\frac{1}{2} \left\{ \frac{\hat{\bm{\pi}}_i}{M}, \hat{\rho}_{i}(\textbf{r}) \right\},
%\end{align*}
\end{eqnarray}
where $\hat{\textbf{p}}_{i}$ and $\hat{\textbf{r}}_{i}$ are the (canonical) momentum and position of particle $i$ respectively, from which 
%$M, \, e$ are the mass and charge of the particles respectively,
$\hat{\bm{\pi}}_{i} \equiv \hat{\textbf{p}}_{i} -e{\textbf{A}}(\hat{\textbf{r}}_{i})$ is its kinetic momentum, and $\{\hat{C},\hat{D}\}\equiv\hat{C}\hat{D} + \hat{D}\hat{C}$.

\begin{figure}[htp]
\centering
\includegraphics[width=3.45in]{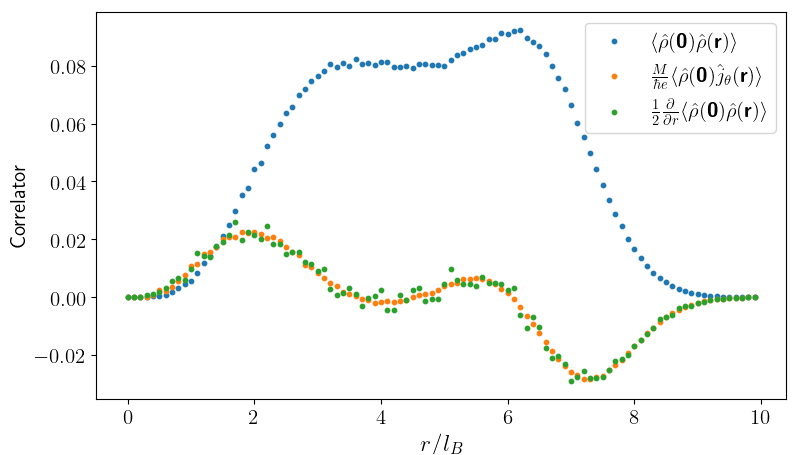}
\caption{ Plot of current density-number density and number density-number density correlators obtained numerically using the Metropolis Monte Carlo method for the filling fraction $\nu = \frac{1}{2}$ Laughlin state in a disc geometry with particle number $N=15$, after $500000$ iterations and discarding the first $80000$ ``thermalisation" runs. The radial derivative of $\langle \hat{\rho}({\textbf 0})\hat{\rho}({\textbf r})\rangle$  can be seen to coincide with the angular component of the current density-number density correlator, $\langle \hat{\rho}({\textbf 0})\hat{j}_\theta({\textbf r})\rangle$, demonstrating the correspondence in Equation (\ref{centralresult}). }
\label{MCLaughlinFigure}
\end{figure}

The key observation which underlies our general result below is that, for states defined in (\ref{LLLstate}), the matrix elements of the current density and number density obey the following relation: 
\begin{equation}
\label{singlebody}
\bra{m_{i}}\hat{j}_{\alpha,i}(\textbf{r})\ket{m^{\prime}_{i}}=\frac{\hbar}{2M} \epsilon_{\alpha\beta} \frac{\partial}{\partial r_{\beta}}\bra{m_{i}}\hat{\rho}_{i}(\textbf{r})\ket{m^{\prime}_{i}}.
\end{equation}
When $m_{i}=m^{\prime}_{i}$, Equation (\ref{singlebody}) can be physically interpreted as the magnetisation current\cite{chr} due to the cyclotron motion, arising from an average magnetisation density in the $z$-direction of $M_{z} = -\frac{\partial E}{\partial B} \braket{\hat{\rho}}$ with the LLL energy $E=\hbar\omega_c/2$. By acting on the one-body states with the current density and number density operators (\ref{densitycurrent}) and using the observation (\ref{singlebody}), the following general result for the many-body LLL state can be derived (see Supplementary Material for further details):
\begin{equation}
\label{centralresult}
\braket{\hat{\rho}(\textbf{r}^{\prime})\hat{j}_{\alpha}(\textbf{r})}_{\rm LLL}=\frac{\hbar}{2M} \epsilon_{\alpha\beta} \frac{\partial}{\partial r_{\beta}}\braket{\hat{\rho}(\textbf{r}^{\prime})\hat{\rho}(\textbf{r})}_{\rm LLL}.
\end{equation}
This result can also be extended to an arbitrary number of current density and number density operator insertions (see Supplementary Material). Since the only requirement for Equation (\ref{centralresult}) to hold is that the $N$-body wavefunction is fully contained in the LLL, it can be applied to various states describing both the integer and fractional quantum Hall effects. In Figure \ref{MCLaughlinFigure}, we show numerically obtained correlators for the bosonic Laughlin state at half filling\cite{MCLaughlin} which are shown to agree with Equation (\ref{centralresult}).

This relationship shows that one cannot extract further information with current density correlators than with number density correlators for LLL states. For single-particle states in higher LLs, Equation (\ref{singlebody}) is no longer valid due to the presence of Laguerre polynomials%[see Equation (\ref{hignnwfs})]
, which introduce non-holomorphic factors to the wavefunctions. However, a result of similar form to (\ref{singlebody}) may be derived for the $n^{\text{th}}$ LL, by replacing the single-particle number density and current density operators with $\hat{\rho}^{(n)}_{i}(\textbf{r})=\frac{1}{n!} (\hat{a}^{\dagger}_{i})^{n}\hat{\rho}_{i}(\textbf{r})(\hat{a}_{i})^{n}$, $\hat{j}^{(n)}_{\alpha,i}(\textbf{r})=\frac{1}{n!} (\hat{a}^{\dagger}_{i})^{n}\hat{j}_{\alpha,i}(\textbf{r})(\hat{a}_{i})^{n}$, where $\hat{a}_{i}$ ($\hat{a}_{i}^{\dagger}$) is the annihilation (creation) operator which lowers (raises) the LL of the $i$-th single particle state, given by 
$\hat{a}_{i} =  (\hat{\pi}_{i,x} + \ii\hat{\pi}_{i,y})l_B/(\sqrt{2}\hbar)$.
Inserting these operators for the many-body correlator reproduces (\ref{centralresult}), so long as the many-body state is fully contained within a single LL and there is no mixing of LLs present. Note that this is no longer a simple correspondence between number density and current density correlators, due to the non-trivial form of the operators $\hat{a}_{i}$, highlighting the special property of the LLL within this context.

When mixing of LLs is present due to interactions, we expect a deviation from the LLL result (\ref{centralresult}). In fact, we will show that this deviation can be used to extract the interaction energy of the system. To explore this connection, we consider a system of bosons with contact interactions: 
\begin{equation}
\label{manybodyHam}
\hat{H}=\frac{1}{2M} \sum_{i=1}^{N} [\hat{\textbf{p}}_{i}-e{\textbf{A}}(\hat{\textbf{r}}_{i})]^{2}  + g \sum_{i<j=1}^{N} \delta(\hat{\textbf{r}}_{i}-\hat{\textbf{r}}_{j}).
\end{equation}
We assume that the contact interaction is weak, $g \ll \hbar \omega_{c} l_{B}^{2}$, so that we may consider the mixing with higher LLs perturbatively. We impose a length-scale cut-off $\lambda_{X}\ll l_B$ for the contact potential, which later plays a role in avoiding divergences in the perturbation theory. As Equation (\ref{centralresult}) only holds in the LLL, we expect deviations due to perturbation theory in higher LLs of the contact interaction. We define the deviation from the LLL result as: 
\begin{equation}
\begin{aligned}
\label{deviation}
\Delta_{\alpha}(\textbf{r})\equiv \epsilon_{\alpha\beta} \frac{\partial}{\partial r_{\beta}}\braket{\hat{\rho}(\textbf{0})\hat{\rho}(\textbf{r})} - \frac{2M}{\hbar}\braket{\hat{\rho}(\textbf{0})\hat{j}_{\alpha}(\textbf{r})}.
\end{aligned}
\end{equation}

We first consider a system of two particles, where analytical expressions for LLL wavefunctions are known. Namely, according to Kohn's theorem\cite{kohn1961cyclotron}, the two-body Hamiltonian can be separated into relative co-ordinate $\hat{\textbf{r}}=\hat{\textbf{r}}_{1}-\hat{\textbf{r}}_{2}$ and centre-of-mass co-ordinate  $\hat{\textbf{R}} = \frac{1}{2}(\hat{\textbf{r}}_{1}+\hat{\textbf{r}}_{2}
)$ terms to obtain an effective one-body Hamiltonian, with interaction potential as a function of the relative co-ordinate only. Perturbation theory can then be performed on LLL wavefunctions of the form given in (\ref{LLLstate}), with effective magnetic length $\Tilde{l}_{B}=\sqrt{2}l_{B}$ i.e. an effective charge $\Tilde{e}=e/2$, due to the introduction of the relative co-ordinate [importantly, this will also be the charge associated with the vector potential in Equation (\ref{densitycurrent})]. 

Note that the centre-of-mass part of the wavefunction is not affected by the contact potential and is not relevant for correlators in a translationally invariant system. Formally, 
we consider a large system, with points $\textbf{r}$ and $\textbf{r}^{\prime}$ deep inside the bulk where
the system is locally translationally invariant. Then the properties can be derived by considering a uniform system, with periodic boundary conditions for which translational invariance is exact. This allows us to integrate out the centre-of-mass contribution in the correlators for a translationally invariant system by using centre-of-mass translation operators $\hat{t}(\textbf{x})=\prod_{i}\exp( -{\rm i} \textbf{x}\cdot\hat{\textbf{K}_{i}})$, where $\hat{\textbf{K}}_{i} = \hat{\bm{\pi}}_{i} - \hbar \textbf{e}_{z} \times \hat{\textbf{r}}_{i} / l_{B}^{2}$ is the ``pseudomomentum" chosen such that the translation operators commute with the dynamical momentum $\hat{\bm{\pi}}_{i}$ \cite{haldane1985many}, and introducing the replacement
\begin{equation}
\label{cheekyreplacement}
\bra{\psi} ... \ket{\psi} \rightarrow \int \bra{\psi} \hat{t}^{\dagger}(\textbf{x})  \ ... \ \hat{t}(\textbf{x})\ket{\psi} d^{2}\textbf{x}
\end{equation}
for all correlators, where $\textbf{x}$ parameterises the translationally invariant system. For example, for the number density correlator, we would have $\braket{\delta(\hat{\textbf{r}}_{1})\delta(\textbf{r}-\hat{\textbf{r}}_{2})} \rightarrow \braket{\delta(\textbf{r}-(\hat{\textbf{r}}_{1}-\hat{\textbf{r}}_{2}))}$. From now on, we will carry out this replacement implicitly and refer back to it when discussing the perturbations of the $N$-body system. 

We use the perturbed wavefunctions to find the first-order deviation (\ref{deviation}) (see Supplementary Material):
\begin{equation}
\label{firstordercorrection}
\begin{aligned}
\Delta_{\theta}(\textbf{r}) = \frac{g}{4 \pi^{2} \hbar \omega_{c} \Tilde{l}_{B}^{6}} r\exp\bigg(-\frac{r^{2}}{2\Tilde{l}_{B}^{2}} \bigg) \sum_{n=1}^{n_{X}} \frac{1}{n} L_{n-1}^{1} \bigg(\frac{r^{2}}{2\Tilde{l}_{B}^{2}}\bigg).
\end{aligned}
\end{equation}
The cut-off in the sum $n_{X}$ ensures that there are no divergences, and corresponds to a minimum range of interaction $\lambda_{X}$ such that $n_{X}\sim (\frac{l_{B}}{\lambda_{X}}
)^{2}$.
Defining the interaction energy as $E_{\text{int}} \equiv \braket{\psi_{0}|\hat{U}_{\text{int}}|\psi_{0}}$, 
the orthogonality relation of generalised Laguerre polynomials (see Supplementary Material\ref{AppB}) can be used to obtain the relation
\begin{equation}
\label{Eint}
E_{\text{int}}=\frac{1}{2} \hbar \omega_{c}  \int |\textbf{r}\times \boldsymbol{\Delta}(\textbf{r})| \, d^{2}\textbf{r}.
\end{equation}
This result shows that the interaction energy of the two-body system can be found by measuring number density and current density correlators that appear in (\ref{deviation}). 

In order to verify the validity of (\ref{Eint}) for a many-body state of the Hamiltonian (\ref{manybodyHam}), we utilise Schmidt decomposition to factorise the $N$-body wavefunction into a system with two particles and $N-2$ particles respectively: 
\begin{equation}
\label{groundstatemanybody}
\Psi^{(0)}(\textbf{r}_{i},\textbf{r}_{j},\{\textbf{r}_{k}\}) = \sum_{m,M} \psi_{m}^{(0)} (\textbf{r}_{ij}) \psi^{(0)}_{M} (2\textbf{R}_{ij}) \Phi^{(0)}_{m,M} (\{\textbf{r}_{k}\}),
\end{equation}
where we have used relative and centre-of-mass angular momentum quantum numbers $m$ and $M$ to label the states, $\psi^{(0)}_{m(M)}$ correspond to single-particle LLL wavefunctions in Equation (\ref{LLLstate}) with angular momentum $m$ ($M$) and effective magnetic length $\Tilde{l}_{B}$,  $\Phi^{(0)}_{m,M}$ is some (unnormalised) wavefunction of $N-2$ particles, and $\textbf{r}_{ij} = \textbf{r}_{i} - \textbf{r}_{j}$, $\textbf{R}_{ij} = (\textbf{r}_{i} + \textbf{r}_{j})/2$ are the relative and centre-of-mass co-ordinates respectively for the positions of particles $i$ and $j$. 

The contact potential then acts pairwise on the many-body wavefunction and perturbatively raises the LL of the 2-body wavefunctions in relative co-ordinate, such that 
\begin{equation}
\label{nstatemanybody}
\Psi^{(n)}(\textbf{r}_{i},\textbf{r}_{j},\{\textbf{r}_{k}\}) = \sum_{m,M} \psi_{m}^{(n)} (\textbf{r}_{ij}) \psi^{(0)}_{M} (2\textbf{R}_{ij}) \Phi^{(0)}_{m,M} (\{\textbf{r}_{k} \}),
\end{equation}
where $\psi_{m}^{(n)}$ is the wavefunction for the $n^{\text{th}}$ LL, and $\{\textbf{r}_{k}\}$ is a set of $N-2$ co-ordinates. By applying (\ref{cheekyreplacement}) to all correlators involving wavefunctions (\ref{groundstatemanybody}) and (\ref{nstatemanybody}) to eliminate two-body centre-of-mass wavefunctions $\psi_{M}^{(0)}$, the perturbation problem for these states can be reduced to that of the two-body case (see Supplementary Material\ref{AppC} for key simplifications). Therefore, we conclude that Equation (\ref{Eint}) can be used to calculate the interaction energy of a system of $N$ interacting particles using current density and number density correlators. 

A useful cross-check of (\ref{Eint}) in the $N$-body context is the Laughlin state of bosons, given by
\begin{equation}
\label{laughlinstate}
\Psi_{L}(z_{1},...,z_{N}) = \prod_{i<j=1}^{N} (z_{i}-z_{j})^{2} \exp\bigg( -\frac{1}{4l_{B}^{2}}\sum_{k=1}^{N} |z_{k}|^{2}  \bigg),
\end{equation}
where $z_{i}=x_{i}+\ii y_{i}$ is the position of the $i^{\text{th}}$ particle in complex form. 
This wavefunction is the ground state of (\ref{manybodyHam}) for filling fraction $\nu=\frac{1}{2}$, with a vanishing interaction energy \cite{wilkin1998attractive}. This is consistent with (\ref{Eint}), as the Laughlin state is fully in the LLL, resulting in a vanishing deviation. We note, however, that other states of (\ref{manybodyHam}) with differing filling fractions, or states at non-zero temperature, will have a non-vanishing interaction energy which could, in turn, be probed by current density correlators as described. 

We have not included a confining potential in the Hamiltonian (\ref{manybodyHam}), as box traps are often used in quantum gas microscopy experiments \cite{navon2021quantum}, whose potential does not impact the bulk correlators of the system. However, if the system of interest has a different confining potential, it would be natural to measure connected correlators of current density, where the operator $\hat{j}_{\alpha}(\textbf{r})$ is replaced by $\hat{j}_{\alpha}(\textbf{r})-\braket{\hat{j}_{\alpha}(\textbf{r})}$. This would remove any one-body contributions to the current arising from the potential~\cite{footnote}.

Finally, we discuss the application of the presented continuum results to lattice systems. In the perturbative calculation, we considered two energy scales: the LL gap $\hbar\omega_{c}$ and the interaction strength $U\equiv{g}/{l_{B}^{2}}$. The lattice introduces an additional energy scale corresponding to the bandwidth of the Chern band $\epsilon$. This induces another contribution to the current density related to the dispersion of the band.  The correlator of current density with number density can then be decomposed into three contributions at the separate energy scales:
\begin{equation}
\braket{\hat{\rho}(\textbf{0})\hat{j}_{\alpha}(\textbf{r})} = \frac{l_{B}}{\hbar} \left[ \hbar\omega_{c}f(\textbf{r})+U g(\textbf{r})+\epsilon h(\textbf{r}) \right].
\end{equation}
The first contribution will have functional scaling $f(\textbf{r})$ related to the number density correlator according to (\ref{centralresult}), whereas $g(\textbf{r})$ corresponds to the deviation given in (\ref{deviation}), and $h(\textbf{r})$ is some further deviation dependent on the energy dispersion of the band. To access a regime of strong interactions where strongly correlated, fractional Chern insulator states can appear, one requires that this bandwidth is small compared to the interaction strength (which is typically small compared to the LL gap) \cite{harper2014perturbative}, i.e. 
\begin{equation}
\epsilon \ll U\ll \hbar\omega_{c} \,.
\end{equation}
Thus, in such a regime, the additional contribution to the current density described by $h(\textbf{r})$ is small, and our continuum results above still provide an accurate description.

\begin{figure} [htp]
\centering  

\subfloat[]{

\tikzset{every picture/.style={line width=0.75pt}} %set default line width to 0.75pt
\begin{tikzpicture}[x=0.75pt,y=0.75pt,yscale=-0.92,xscale=0.92]
%Shape: Grid [id:dp22478906055530967] 
\draw  [draw opacity=0] (120,44) -- (383,44) -- (383,212) -- (120,212) -- cycle ; \draw   (147,44) -- (147,212)(200,44) -- (200,212)(253,44) -- (253,212)(306,44) -- (306,212)(359,44) -- (359,212) ; \draw   (120,71) -- (383,71)(120,124) -- (383,124)(120,177) -- (383,177) ; \draw    ;

%Shape: Ellipse [id:dp03266800632523359] 
\draw  [fill={rgb, 255:red, 0; green, 0; blue, 0 }  ,fill opacity=1 ] (248,124) .. controls (248,121.24) and (250.24,119) .. (253,119) .. controls (255.76,119) and (258,121.24) .. (258,124) .. controls (258,126.76) and (255.76,129) .. (253,129) .. controls (250.24,129) and (248,126.76) .. (248,124) -- cycle ;
%Shape: Ellipse [id:dp15097538520771614] 
\draw  [fill={rgb, 255:red, 74; green, 144; blue, 226 }  ,fill opacity=1 ] (258,71) .. controls (258,68.24) and (255.76,66) .. (253,66) .. controls (250.24,66) and (248,68.24) .. (248,71) .. controls (248,73.76) and (250.24,76) .. (253,76) .. controls (255.76,76) and (258,73.76) .. (258,71) -- cycle ;
%Shape: Ellipse [id:dp0891031231657532] 
\draw  [fill={rgb, 255:red, 74; green, 144; blue, 226 }  ,fill opacity=1 ] (195,124) .. controls (195,121.24) and (197.24,119) .. (200,119) .. controls (202.76,119) and (205,121.24) .. (205,124) .. controls (205,126.76) and (202.76,129) .. (200,129) .. controls (197.24,129) and (195,126.76) .. (195,124) -- cycle ;
%Shape: Ellipse [id:dp8607984415819442] 
\draw  [fill={rgb, 255:red, 74; green, 144; blue, 226 }  ,fill opacity=1 ] (301,124) .. controls (301,121.24) and (303.24,119) .. (306,119) .. controls (308.76,119) and (311,121.24) .. (311,124) .. controls (311,126.76) and (308.76,129) .. (306,129) .. controls (303.24,129) and (301,126.76) .. (301,124) -- cycle ;
%Shape: Ellipse [id:dp5994217956934448] 
\draw  [fill={rgb, 255:red, 74; green, 144; blue, 226 }  ,fill opacity=1 ] (248,177) .. controls (248,174.24) and (250.24,172) .. (253,172) .. controls (255.76,172) and (258,174.24) .. (258,177) .. controls (258,179.76) and (255.76,182) .. (253,182) .. controls (250.24,182) and (248,179.76) .. (248,177) -- cycle ;
%Straight Lines [id:da0718154691693842] 
\draw [color={rgb, 255:red, 208; green, 2; blue, 27 }  ,draw opacity=1 ][fill={rgb, 255:red, 208; green, 2; blue, 27 }  ,fill opacity=1 ]   (306,119) -- (306,71) ;
\draw [shift={(306,88)}, rotate = 90] [fill={rgb, 255:red, 208; green, 2; blue, 27 }  ,fill opacity=1 ][line width=0.08]  [draw opacity=0] (12,-3) -- (0,0) -- (12,3) -- cycle    ;
%Straight Lines [id:da49086732333411853] 
\draw [color={rgb, 255:red, 208; green, 2; blue, 27 }  ,draw opacity=1 ]   (306,71) -- (258,71) ;
\draw [shift={(275,71)}, rotate = 360] [fill={rgb, 255:red, 208; green, 2; blue, 27 }  ,fill opacity=1 ][line width=0.08]  [draw opacity=0] (12,-3) -- (0,0) -- (12,3) -- cycle    ;
%Straight Lines [id:da8102527607455463] 
\draw [color={rgb, 255:red, 208; green, 2; blue, 27 }  ,draw opacity=1 ]   (248,71) -- (200,71) ;
\draw [shift={(217,71)}, rotate = 360] [fill={rgb, 255:red, 208; green, 2; blue, 27 }  ,fill opacity=1 ][line width=0.08]  [draw opacity=0] (12,-3) -- (0,0) -- (12,3) -- cycle    ;
%Straight Lines [id:da7501041544948319] 
\draw [color={rgb, 255:red, 208; green, 2; blue, 27 }  ,draw opacity=1 ]   (200,71) -- (200,119) ;
\draw [shift={(200,102)}, rotate = 270] [fill={rgb, 255:red, 208; green, 2; blue, 27 }  ,fill opacity=1 ][line width=0.08]  [draw opacity=0] (12,-3) -- (0,0) -- (12,3) -- cycle    ;
%Straight Lines [id:da22844379399381076] 
\draw [color={rgb, 255:red, 208; green, 2; blue, 27 }  ,draw opacity=1 ]   (200,129) -- (200,177) ;
\draw [shift={(200,160)}, rotate = 270] [fill={rgb, 255:red, 208; green, 2; blue, 27 }  ,fill opacity=1 ][line width=0.08]  [draw opacity=0] (12,-3) -- (0,0) -- (12,3) -- cycle    ;
%Straight Lines [id:da4142083486052298] 
\draw [color={rgb, 255:red, 208; green, 2; blue, 27 }  ,draw opacity=1 ]   (200,177) -- (248,177) ;
\draw [shift={(231,177)}, rotate = 180] [fill={rgb, 255:red, 208; green, 2; blue, 27 }  ,fill opacity=1 ][line width=0.08]  [draw opacity=0] (12,-3) -- (0,0) -- (12,3) -- cycle    ;
%Straight Lines [id:da626846970602331] 
\draw [color={rgb, 255:red, 208; green, 2; blue, 27 }  ,draw opacity=1 ]   (258,177) -- (306,177) ;
\draw [shift={(289,177)}, rotate = 180] [fill={rgb, 255:red, 208; green, 2; blue, 27 }  ,fill opacity=1 ][line width=0.08]  [draw opacity=0] (12,-3) -- (0,0) -- (12,3) -- cycle    ;
%Straight Lines [id:da7449067563502224] 
\draw [color={rgb, 255:red, 208; green, 2; blue, 27 }  ,draw opacity=1 ]   (306,177) -- (306,129) ;
\draw [shift={(306,146)}, rotate = 90] [fill={rgb, 255:red, 208; green, 2; blue, 27 }  ,fill opacity=1 ][line width=0.08]  [draw opacity=0] (12,-3) -- (0,0) -- (12,3) -- cycle    ;

%Straight Lines [id:da2633224621221154] 
\draw    (254,228) -- (304,228) ;
\draw [shift={(306,228)}, rotate = 180] [color={rgb, 255:red, 0; green, 0; blue, 0 }  ][line width=0.75]    (10.93,-3.29) .. controls (6.95,-1.4) and (3.31,-0.3) .. (0,0) .. controls (3.31,0.3) and (6.95,1.4) .. (10.93,3.29)   ;
%Straight Lines [id:da9151095874098708] 
\draw    (306,228) -- (256,228) ;
\draw [shift={(254,228)}, rotate = 360] [color={rgb, 255:red, 0; green, 0; blue, 0 }  ][line width=0.75]    (10.93,-3.29) .. controls (6.95,-1.4) and (3.31,-0.3) .. (0,0) .. controls (3.31,0.3) and (6.95,1.4) .. (10.93,3.29)   ;

% Text Node
\draw (275,212.4) node [anchor=north west][inner sep=0.75pt]    {$a$};
\draw (258,108.4) node [anchor=north west][inner sep=0.75pt]    {$(x_{c},y_{c})$};
\draw (310,108.4) node [anchor=north west][inner sep=0.75pt]    {$(i)$};
\draw (310,146.4) node [anchor=north west][inner sep=0.75pt]    {$(ii)$};

\end{tikzpicture}}

\subfloat[]{\includegraphics[width=.4\textwidth]{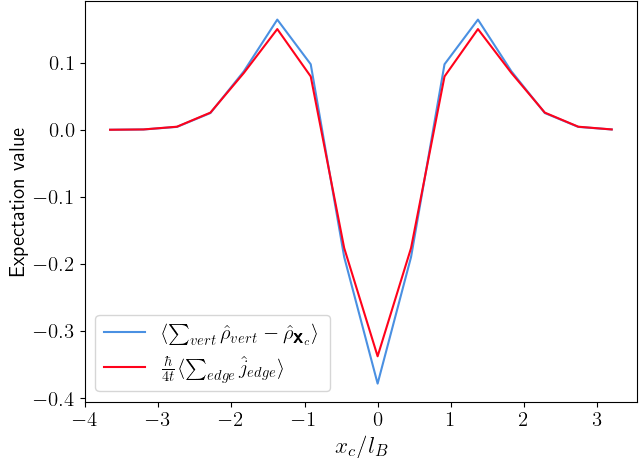}}

\caption{(a) Schematic diagram of the discrete density and current density  operators for the central site at $(x_{c},y_{c})$. The blue circles $(i)$ correspond to number density operators defined on vertices, whereas the red arrows $(ii)$ correspond to current density operators on the edges. (b) Graph of the expectation values given in (\ref{discretecentralresult}) for the density and current operators respectively, acting on the wavefunction (\ref{LandauGaugeWF}) centred at the origin ($k_{y}=0$). We have taken $n_{\phi}=\frac{1}{30}$, such that the lattice spacing is $a = \sqrt{\frac{2 \pi}{30}} l_{B}$. In the limit of smaller magnetic flux, the difference between the two curves approaches zero. }
\label{latticefigs}
\end{figure}
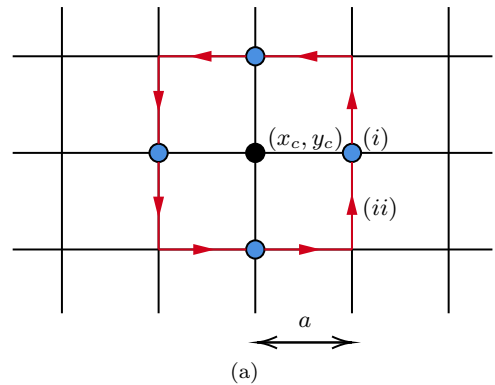
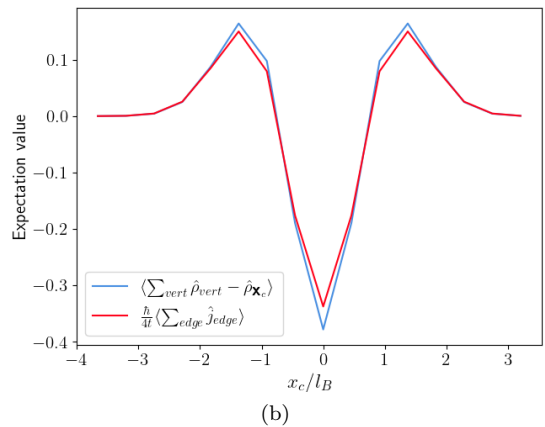

The simplest regime to demonstrate the applicability of our results to the lattice is the Harper-Hofstadter model in the low magnetic flux limit. There, the ground states are given exactly by the continuum LLL wavefunctions, as the length scale $l_{B}$ becomes larger than the lattice spacing $a$. The Harper-Hofstadter Hamiltonian \cite{hofstadter1976energy, harper1955general} is given by
\begin{equation}
\label{harperham}
\hat{H} = - t \sum_{\textbf{x}} \hat{c}^{\dagger}_{\textbf{x}+a \textbf{e}_{x}}\hat{c}_{\textbf{x}} + \ee^{\ii2\pi n_{\phi} x}\hat{c}^{\dagger}_{\textbf{x}+a\textbf{e}_{y}} \hat{c}_{\textbf{x}} + \text{h.c.} ,
\end{equation}
where $t$ is the hopping amplitude, $\textbf{x}$ are the co-ordinates of sites on the lattice, $\textbf{e}_{x}$, $\textbf{e}_{y}$ are the unit vectors in the $x$ and $y$ directions respectively, and $n_{\phi} = {p}/{q}$ is the magnetic flux per unit cell (where $p$ and $q$ are integers for a system with periodic boundary conditions). We note that we have now switched to the Landau gauge, $\textbf{A}(\textbf{r})=Bx \textbf{e}_{y}$, such that the LLL wavefunctions are given by
\begin{equation}
\braket{\textbf{r}|k_{y}} = \mathcal{N} \ee^{\ii k_{y} y} \exp\bigg[-\frac{(x-k_{y}l_{B}^{2})^{2}}{2l_{B}^{2}}\bigg],
\label{LandauGaugeWF}
\end{equation}
where $\mathcal{N}$ is the normalisation of the wavefunctions on the lattice, and the states are labeled by the momentum quantum number $k_{y}$. 

To apply the continuum results to the lattice, we take the curl of both sides of Equation (\ref{singlebody}) and then discretise the Laplacian acting on the density matrix elements. Noting that, on the lattice, the current density operator is defined on bonds, we approximate the curl of the current density matrix element by a line integral around a closed loop surrounding four unit cells [see Figure \ref{latticefigs} (a)] divided by the area of the enclosed region, giving the following discretisation: 
\begin{equation}
\label{discretecentralresult}
\bra{k_{y}}\frac{1}{4}\sum_{\text{edge}} \hat{j}_{\text{edge}} \ket{k_{y}^{\prime}} \approx \frac{t}{\hbar} \bra{k_{y}}\sum_{\text{vert}} \hat{\rho}_{\text{vert}} - 4 \hat{\rho}_{\textbf{x}_{c}} \ket{k_{y}^{\prime}}
\end{equation}
where $a$ is the lattice spacing, the two sums are over edges and vertices as depicted in Figure \ref{latticefigs} (a) and centred at the point $\textbf{x}_{c}=(x_{c},y_{c})$, and the operators on the lattice are now given by
$\hat{\rho}_{\textbf{x}} = \hat{c}^{\dagger}_{\textbf{x}}\hat{c}_{\textbf{x}} 
$,
$\hat{j}_{\textbf{x}\rightarrow\textbf{x} + a \textbf{e}_{x}} = -\frac{\ii t}{\hbar}  \hat{c}^{\dag}_{\textbf{x}}\hat{c}_{\textbf{x}+a \textbf{e}_{x}} + \text{h.c.}
$, $\hat{j}_{\textbf{x}\rightarrow\textbf{x} + a \textbf{e}_{y}} = -\frac{\ii t}{\hbar} \ee^{\ii 2\pi n_{\phi}x} \hat{c}^{\dag}_{\textbf{x}}\hat{c}_{\textbf{x}+a \textbf{e}_{y}} + \text{h.c.}
$ and $\hat{c}_{\textbf{x}}$ is the particle annihilation operator on the lattice. We have also performed the replacement $\frac{\hbar}{2 M a^{2}}\rightarrow \frac{t}{\hbar}$ to obtain the lattice version of the continuum formula ($\ref{singlebody}$). Figure \ref{latticefigs} (b) demonstrates the approximate correspondence between expectation values given in (\ref{discretecentralresult}), where $k_{y}=k_{y}^{\prime}$. This is to be expected, as the lattice operators are equivalent to the finite difference approximations of the continuum operators given in (\ref{densitycurrent}).

To conclude, we have presented a direct relationship between current density and number density correlators for many-body states in the LLL, which does not have a simple correspondence in higher LLs. Furthermore, we have shown that the deviations from this relationship can be used to determine the system's interaction energy for states which are perturbatively lifted out of the LLL via interactions. We have also discussed the applicability of our results to lattice settings.  
The natural continuation of this work is to investigate the properties of current density correlators for many-body states on lattice models, such as the Harper-Hofstadter model.
%extend the findings in the continuum to the Harper-Hofstadter model away from the low magnetic flux limit, as this is the relevant regime for experimental realisations of many-body FQH and fractional Chern insulator states. 
Furthermore, it would be interesting to explore connections between correlators and analogous quantities in the corresponding field theories (e.g. Chern-Simons theory), where the emergent gauge field couples the number density of particles with the current density \cite{giuliani2008quantum}.
%, it may be possible to relate the current density correlators in the field theoretic approach with the results derived in this paper. 

\textit{Acknowledgements -- }  
This work was supported by the Engineering and Physical Sciences Research Council [grant number EP/V062654/1], a Simons Investigator Award [Grant No. 511029] and a Cambridge International Scholarship provided by the Cambridge Trust. 
For the purpose of open access, the authors have applied a creative commons attribution (CC BY) licence to any author accepted manuscript version arising.

\nocite{*}

\bibliography{library.bib}% Produces the bibliography via BibTeX.

\appendix

\section{Supplementary Material}
\subsection{Further Details of Theory in LLL}
\label{AppA}

By applying operators in Equation (\ref{densitycurrent}) on the state (\ref{LLLmainstate}), we obtain the following expressions for the correlators:
\begin{equation}
\begin{aligned}
\label{appafirsteq}
\braket{\hat{\rho}(\textbf{r}^{\prime})\hat{\textbf{j}}(\textbf{r})}_{\rm LLL}=  \sum_{\{m_{i}\}}\sum_{\{m^{\prime}_{j}\}} c^{*}_{m_{1}...m_{N}}c_{m^{\prime}_{1}..m^{\prime}_{N}} & \\ \times \sum_{k=1}^{N}\sum_{l=1}^{N} \bra{m_{k}}\bra{m_{l}}\hat{\rho}_{k}(\textbf{r}^{\prime})\hat{\textbf{j}}_{l}(\textbf{r}) \ket{m^{\prime}_{l}}\ket{m^{\prime}_{k}} 
\\
=  \sum_{\{m_{i}\}}\sum_{\{m^{\prime}_{j}\}} c^{*}_{m_{1}...m_{N}}c_{m^{\prime}_{1}..m^{\prime}_{N}} & \\ \times \Biggl(\sum_{k\neq l} \bra{m_{k}}\hat{\rho}_{k}(\textbf{r}^{\prime})\ket{m^{\prime}_{k}}\bra{m_{l}}\hat{\textbf{j}}_{l}(\textbf{r}) \ket{m^{\prime}_{l}}
\\ + \sum_{k=1}^{N}\bra{m_{k}}\bra{m_{k}}\hat{\rho}_{k}(\textbf{r}^{\prime})\hat{\textbf{j}}_{k}(\textbf{r}) \ket{m^{\prime}_{k}}\ket{m^{\prime}_{k}}\Biggl),
\end{aligned}
\end{equation}
where we have split the sum over $k,l$ into two parts: the first part corresponds to number density and current density operators acting on two separate single-particle states, and the second corresponds to the operators acting on the same state. 

Using the fact that $\delta(\textbf{r}-\hat{\textbf{r}}_{i}) = \ket{\textbf{r}}\bra{\textbf{r}}_{i}$, it can be shown that the second sum in (\ref{appafirsteq}) is proportional to $\delta(\textbf{r}-\textbf{r}^{\prime})$. Hence, this contribution to the correlator vanishes by requiring that $\textbf{r}\neq\textbf{r}^{\prime}$. The current density correlator can then be re-written as

\begin{equation}
\label{appathird}
\begin{aligned}
\braket{\hat{\rho}(\textbf{r}^{\prime})\hat{\textbf{j}}(\textbf{r})}_{\rm LLL}=  \sum_{\{m_{i}\}}\sum_{\{m^{\prime}_{j}\}} c^{*}_{m_{1}...m_{N}}c_{m^{\prime}_{1}..m^{\prime}_{N}} & \\ \times \sum_{k\neq l} \braket{m_{k}|\textbf{r}^{\prime}}\braket{\textbf{r}^{\prime}|m^{\prime}_{k}} \bra{m_{l}}\hat{\textbf{j}}_{l}(\textbf{r})\ket{m^{\prime}_{l}}.
\end{aligned}
\end{equation}
To calculate the one-body current density matrix elements, we use the wavefunctions in (\ref{LLLstate}) and the current density operator defined in (\ref{densitycurrent}) in polar co-ordinates. Doing so, we obtain
\begin{equation}
\label{appasecond}
\begin{aligned}
\bra{m_{l}}\hat{j}_{l,r}(\textbf{r})\ket{m_{l}^{\prime}}=\frac{\hbar}{M}\left(\frac{{\rm i}(m_{l}-m^{\prime}_{l})}{2r}\right) \braket{m_{l}|\textbf{r}}\braket{\textbf{r}|m^{\prime}_{l}}, \\ 
\bra{m_{l}}\hat{j}_{l,\theta}(\textbf{r})\ket{m_{l}^{\prime}}=\frac{\hbar}{M}\left(\frac{m_{l}+m^{\prime}_{l}}{2r}-\frac{r}{2l_{B}^{2}}\right) \braket{m_{l}|\textbf{r}}\braket{\textbf{r}|m^{\prime}_{l}},
\end{aligned}
\end{equation}
where $\hat{j}_{l,r}$, $\hat{j}_{l,\theta}$ are the radial and angular components of the current density operator on state $l$ respectively. 
The key observation made now is that the density matrix elements can be related to (\ref{appasecond}) in a precise way given in Equation (\ref{singlebody}). Namely, we have that
\begin{equation}
\begin{aligned}
\frac{1}{r}\frac{\partial}{\partial \theta}\bra{m_{l}}\hat{\rho}_{l}(\textbf{r})\ket{m_{l}^{\prime}}=\left(\frac{{\rm i}(m_{l}^{\prime}-m_{l})}{r}\right) \braket{m_{l}|\textbf{r}}\braket{\textbf{r}|m^{\prime}_{l}}, \\ 
\frac{\partial}{\partial r} \bra{m_{l}}\hat{\rho}_{l}(\textbf{r})\ket{m_{l}^{\prime}}=\left(\frac{m_{l}+m^{\prime}_{l}}{r}-\frac{r}{l_{B}^{2}}\right) \braket{m_{l}|\textbf{r}}\braket{\textbf{r}|m^{\prime}_{l}},
\end{aligned}
\end{equation}
from which (\ref{singlebody}) follows. Substituting (\ref{singlebody}) into (\ref{appathird}), we obtain
\begin{equation}
\label{appafourth}
\begin{aligned}
\braket{\hat{\rho}(\textbf{r}^{\prime})\hat{j}_{\alpha}(\textbf{r})}_{\rm LLL}=  \frac{\hbar}{2M} \epsilon_{\alpha\beta} \frac{\partial}{\partial r_{\beta}}\sum_{\{m_{i}\}}\sum_{\{m^{\prime}_{j}\}} c^{*}_{m_{1}...m_{N}}c_{m^{\prime}_{1}..m^{\prime}_{N}} & \\ \times \sum_{k\neq l} \braket{m_{k}|\textbf{r}^{\prime}}\braket{\textbf{r}^{\prime}|m^{\prime}_{k}} \braket{m_{l}|\textbf{r}}\braket{\textbf{r}|m^{\prime}_{l}}.
\end{aligned}
\end{equation}
By noticing that the expression on the right-hand side after the derivative is just the number density correlator $\braket{\hat{\rho}(\textbf{r}^{\prime})\hat{\rho}(\textbf{r})}_{\rm LLL}$, the expression reduces to (\ref{centralresult}). 

We note that the above derivation can be carried out with arbitrarily many additional insertions of number density and current density operators, from which we get the general result
\begin{align*}
\braket{\prod_{a=1}^{A}\hat{\rho}(\textbf{r}^{(a)})\prod_{b=1}^{B}\hat{j}_{\alpha_{b}}(\textbf{r}^{(b)})}_{\rm LLL}=\left(\frac{\hbar e}{2M}\right)^{B} \prod_{b=1}^{B} \epsilon_{\alpha_{b}\beta_{b}} \frac{\partial}{\partial\textbf{r}^{(b)}_{\beta_{b}}}\\
\times\braket{\prod_{a=1}^{A}\hat{\rho}(\textbf{r}^{(a)})\prod_{b=1}^{B}\hat{\rho}(\textbf{r}^{(b)})}_{\rm LLL},
\end{align*}
where $a$ ($b$) is the index enumerating each of $A$ ($B$) number density (current density) operators in the correlator on the LHS, $\alpha_{b},\beta_{b}$ are the co-ordinate indices for the $b^{\text{th}}$ operator, and we require as before that $\textbf{r}^{(a)}\neq \textbf{r}^{(b)}$ for all index values.

\subsection{Perturbation of Two Particles}
\label{AppB}
The Hamiltonian (\ref{manybodyHam}) for $N=2$ can be written as the sum $\hat{H}=\hat{H}_{0}(\hat{\textbf{r}})+\hat{H}_{0}(\hat{\textbf{R}})+ \hat{U}_{\text{int}}(\hat{\textbf{r}})$ in relative and centre-of-mass co-ordinate operators $\hat{\textbf{r}}$ and $\hat{\textbf{R}}$ respectively. The Hamiltonian $\hat{H}_{0}$ corresponds to a particle in a perpendicular magnetic field whose solutions correspond to wavefunctions given in Equation ($\ref{LLLstate}$), where the quantum numbers $m\ (M)$ are used for the relative (centre-of-mass) co-ordinate. For a bosonic state, $m$ must be even to satisfy the required exchange statistics. 

Performing first-order perturbation theory on the relative co-ordinate wavefunctions, we find that 
\begin{equation}
\label{perturb}
\delta\psi(\textbf{r}) = -\frac{g}{2\pi\hbar \omega_{c}\Tilde{l}_{B}^{2}}\sum_{n=1}^{n_{X}} \frac{1}{ n} \psi_{0}^{(n)} (\textbf{r}) ,
\end{equation}
where $g$ is the strength of the contact potential, $\Tilde{l}_{B}=\sqrt{2}l_{B}$, $n_{X}$ is the cut-off imposed to avoid  divergences, and $\psi_{m}^{(n)}$ is the wavefunction for the $n^{\text{th}}$ Landau level given by 
\begin{equation}
\begin{aligned}
\label{hignnwfs}
\psi_{m}^{(n)} (\textbf{r}) = \sqrt{\frac{n!}{2\pi l_{B}^{2} (n+m)!}} \left( \frac{r}{\sqrt{2}l_{B}}  \right)^{m} \exp\left( -\frac{r^{2}}{4l_{B}^{2}} \right)&  \\ \times L^{m}_{n}\left( \frac{r^{2}}{2l_{B}^{2}}  \right)\ee^{{\rm i}m\theta},
\end{aligned}
\end{equation}
where $L_{n}^{m}$ are the generalised Laguerre polynomials. Note that in (\ref{perturb}), the magnetic length in the wavefunctions above is the effective magnetic length $\Tilde{l}_{B}$. To calculate the first order deviation (\ref{deviation}) beyond the LLL result, we thus have to compute $\bra{\delta\psi(\textbf{r}_{1}-\textbf{r}_{2})} \hat{\rho} (\textbf{0})\hat{\rho}(\textbf{r})\ket{\psi(\textbf{r}_{1}-\textbf{r}_{2})}$ and $\bra{\delta\psi(\textbf{r}_{1}-\textbf{r}_{2})} \hat{\rho} (\textbf{0})\hat{\textbf{j}}(\textbf{r})\ket{\psi(\textbf{r}_{1}-\textbf{r}_{2})}$. Doing so, one obtains Equation (\ref{firstordercorrection}). 

In order to derive Equation (\ref{Eint}), we use the orthogonality relation of generalised Laguerre polynomials given by
\begin{equation}
\int_{0}^{\infty} L_{n}^{k}(x) L_{n^{\prime}}^{k}(x) x^{k} \ee^{-x} dx = \frac{(n+k)!}{n!} \delta_{n,n^{\prime}},
\end{equation}
and the fact that $L_{0}^{k}(x) = 1$, to eliminate all terms except for $i=1$ and obtain

\begin{equation}
\int_{0}^{\infty} \Delta_{\theta}(r) r^{2} dr = \frac{g}{2\pi^{2}\hbar\omega_{c} \Tilde{l}_{B}^{2}},
\end{equation}
which reduces to (\ref{Eint}) when we substitute $E_{\text{int}} = \frac{g}{2\pi\Tilde{l}_{B}^{2}}$. 

\subsection{Perturbation of $N$ Particles}
\label{AppC}
To generalise the two-body perturbative calculation to the $N$-body system, 
%in Appendix \ref{AppB}, 
we use the Schmidt decomposition that separates the many body state into a two-particle state, which is perturbatively raised to higher LLs, and an $N-2$ particle state. This allows us to write a first-order correction to the ground state wavefunction (\ref{groundstatemanybody})

\begin{equation}
\begin{aligned}
\delta \Psi(\textbf{r}_{k},\textbf{r}_{l}, \{\textbf{r}_{s}\}) = - \frac{g}{\hbar\omega_{c} }\sum_{k<l}\sum_{n} \frac{\braket{\Psi^{(n)}|\delta(\textbf{r}_{k}-\textbf{r}_{l})|\Psi^{(0)}}}{ n} \\ \times\Psi^{(n)}(\textbf{r}_{k},\textbf{r}_{l}, \{\textbf{r}_{s}\}) \\ = - \frac{g}{2\pi\hbar\omega_{c} \Tilde{l}_{B}^{2}}\sum_{k<l}\sum_{n}\sum_{M} \frac{\braket{\Phi^{(0)}_{0,M}|\Phi^{(0)}_{0,M}}}{ n} \\ \times\Psi^{(n)}(\textbf{r}_{k},\textbf{r}_{l},\{\textbf{r}_{s}\}),
\end{aligned}
\end{equation}
where $\Phi^{(0)}_{0,M}$ are (unnormalised) wavefunctions of $N-2$ particles. In the rest of this section, we will drop $\{\textbf{r}_{s}\}$ for brevity. 

Writing the number density and current density correlators as $\bra{\delta\Psi_{k,l}} \hat{\rho}_{i} (\textbf{0})\hat{\rho}_{j}(\textbf{r})\ket{\Psi}$ and $\bra{\delta\Psi_{k,l}} \hat{\rho}_{i} (\textbf{0})\hat{\textbf{j}}_{j}(\textbf{r})\ket{\Psi}$ (where there is an implicit sum over pairs of $i,j$ and $k,l$ respectively), it becomes clear that there are three distinct cases: 
\begin{enumerate}[(i)]
\item The trivial case is one where $i,j\neq k,l$, where neither the density nor current density operators act on the pair of coordinates that have been perturbatively excited to higher LLs. Because of this, the number density and current density operators act purely on the $N-2$ body wavefunction which is in the LLL. Therefore, the correlators obey the LLL result (\ref{centralresult}) and there is no correction. 
\item The second case is one for which $i\neq k$, but $j=l$. In this case, the number density operator $\hat{\rho}_{i}(\textbf{0})$ acts on the $N-2$ particle wavefunction, whereas the operator $\hat{\rho}_{l}(\textbf{r})$ (or $\hat{\textbf{j}}_{j}(\textbf{r})$) acts on the excited two body state. Considering first the number density correlator, and using the replacement (\ref{cheekyreplacement}), we have that

\begin{equation}
\begin{aligned}
\bra{\delta\Psi_{k,l}} \hat{\rho}_{i} (\textbf{0})\hat{\rho}_{j}(\textbf{r})\ket{\Psi}  = \sum_{M}\int\prod_{s}  d^{2}\textbf{r}_{s}  |\Phi_{0,M}^{(0)}(0,\{\textbf{r}_{s}\})|^{2} \\ \times \int  d^{2}\textbf{r}_{k} (\psi^{(n)}_{0}(\textbf{r}_{k}-\textbf{r}))^{*} \psi^{(0)}_{0}(\textbf{r}_{k}-\textbf{r}) ,
\end{aligned}
\end{equation}
where we have performed the Schmidt decomposition of the ground state (\ref{groundstatemanybody}). Due to the orthogonality of wavefunctions given in (\ref{hignnwfs}), the following contribution to the correlator vanishes. Using similar arguments, it can be shown that the contribution due to the current density also vanishes. 
\item The third case is one for which $i=k$ and $j=l$ i.e. where the two-point correlator acts directly on the excited pair of particles. This case recovers the results obtained in the two-body section, where the interaction energy can be found to be $E_{\text{int}} = \frac{N(N-1)}{2} \frac{g}{2\pi\Tilde{l}_{B}^{2}}\sum_{M} \braket{\Phi^{(0)}_{0,M}|\Phi^{(0)}_{0,M}} $. Note that the calculation of the norm of the $N-2$ particle wavefunctions drops out as it is present both in the expressions for the deviation and the interaction energy. 
\end{enumerate}

\subsection{Current Density Corrections in a Harmonic Trap}
\label{AppD}

Consider a harmonic trap with confining potential of the form $V(r) = \frac{1}{2}\alpha r^{2}$. For a $\nu=1/2$ FQH state of $N$ particles with a maximum radius $R\approx\sqrt{N}l_{B}$, one often requires that the potential at $R$ be less than the strength of interactions so that it is not energetically favourable for particles to move from the edge of the sample to the bulk, thereby creating excitations. Therefore, one requires the following inequality to hold:
\begin{equation}
V(R) \leq \frac{g}{l_{B}^{2}}\equiv U.
\end{equation}
Since the velocity of excitations due to the confining potential is proportional to the electric field, we have that
\begin{equation}
\nabla V  = \alpha R \leq \frac{2 U}{R},
\end{equation}
where $\nabla V\equiv \frac{\partial}{\partial r} V(r)|_{r=R}$. Hence, the energy scale of the current density correction due to the confining potential is upper bounded by
\begin{equation}
l_{B}\nabla V \leq \frac{2U}{\sqrt{N}}, 
\end{equation}
such that the contribution to current density correlators due to perturbative mixing from the harmonic trap decreases compared to that of interactions as the number of particles $N$ increases, so long as we are working in the regime outlined above. 

\end{document}